\documentclass[12pt,preprint]{aastex}
\def\AU{{\rm AU}}

\begin{document}

\title{The Scattered Disk as the source of the Jupiter Family comets}

\author{Kathryn Volk, Renu Malhotra}
\affil{Lunar and Planetary Laboratory, University of Arizona, Tucson, AZ 85721, USA.}

\begin{abstract}

The short period Jupiter family comets (JFCs) are thought to originate in the Kuiper Belt; specifically, a dynamical subclass of the Kuiper Belt known as the `scattered disk' is argued to be the dominant source of JFCs.  However, the best estimates from observational surveys indicate that this source may fall short by more than two orders of magnitude the estimates obtained from theoretical models of the dynamical evolution of Kuiper belt objects into JFCs. We re-examine the scattered disk as a source of the JFCs and make a rigorous estimate of the discrepancy. We find that the uncertainties in the dynamical models combined with a change in the size distribution function of the scattered disk at faint magnitudes (small sizes) beyond the current observational limit offer a possible but problematic resolution to the discrepancy. We discuss several other possibilities: that the present population of JFCs is a large fluctuation above their long term average, that larger scattered disk objects tidally break-up into multiple fragments during close planetary encounters as their orbits evolve from the trans-Neptune zone to near Jupiter, or that there are alternative source populations that contribute significantly to the JFCs.  Well-characterized observational investigations of the Centaurs, objects that are transitioning between the trans-Neptune Kuiper belt region and the inner solar system, can test the predictions of the non-steady state and the tidal break-up hypotheses.  The classical and resonant classes of the Kuiper belt are worth re-consideration as significant additional or alternate sources of the JFCs. 

\end{abstract}


\section{Introduction}\label{s:Introduction}

It is currently widely accepted that the Jupiter family short period comets (JFCs) and the Centaurs are objects that have escaped from the Kuiper Belt. Studies of the dynamics of Kuiper Belt objects (KBOs) have identified dynamical pathways by which the Kuiper Belt could serve as a source reservoir for the JFCs.  Some of these studies have also made quantitative theoretical predictions of the population of KBOs required to supply the observed population of JFCs in steady state~\citep{holman93, duncan95, levison97, duncan97, morby97, emel04, fernandez04}.  Traditional observational surveys to test these predictions are exceedingly challenging due to the small (sub--10 km) sizes of known JFCs, hence faint magnitudes, $m\gtrsim29$, for their Kuiper Belt precursors.  The use of stellar occultations to detect very small KBOs might solve this problem; to date there have been no universally accepted detections using this method, but \citet{bickerton08} use the lack of detections in the stellar occultation data that has been analyzed so far to place an upper limit on the total number of KBOs larger than $D \sim 1$ km of $~3 \times 10^{9}$ $deg^{-2}$.  It might also be possible to use distortions in the cosmic microwave background or measurements of the gamma-ray flux from cosmic-rays interacting with outer solar system material to constrain the mass distribution in the trans-Neptunian region \citep{babich07, moskalenko08}, but these methods have not yet been implemented.

Several large ground-based KBO surveys have found that the brightness distribution of KBOs brighter than about $m\simeq23$ are described well by a single power law function~\citep{trujillo01cfht,sheppard00,jewitt98}.  Upon extrapolating this function to fainter magnitudes, the theory of the JFC--KB link seemed to be secure.  Indeed, according to the extrapolations, several dynamical classes within the Kuiper Belt -- the classical KBOs, the Plutinos, or the scattered disk -- could potentially be viable sources of the JFCs.  In a detailed review of the problem, \citet{duncan04} concluded that the scattered disk was likely the dominant source of the JFCs, based on then--current theoretical models and observations.  This conclusion has implications for not only the orbital evolution of SDOs, Centaurs, and JFCs, but also for their physical evolution; knowing the specific source region for the JFCs and Centaurs would perhaps allow for a better interpretation of their physical properties such as color and albedo.  Due to their extremely large eccentricities and large semi-major axes that carry them out past the solar wind's termination shock (located at $\sim 85$ AU) and into the local interstellar medium, SDOs experience a much different radiation history than their classical belt counterparts.  \citet{cooper03} discuss the effect of these different radiation environments combined with several resurfacing processes on the colors of the various dynamical classes of KBOs,  finding some evidence that the classical belt is best situated for producing red objects, while other classes would be more neutrally colored.  It is interesting to note that \citet{tegler08} find a bimodal color distribution for the observed Centaurs, and conclude that two different source populations could be a possible explanation; identification of the source population(s) for the Centaurs and the JFCs puts these observations into useful context.
 
Since the review of \citet{duncan04} that identified the SDOs as the most likely source region for the JFCs, there have been two developments that motivate a re-examination of this conclusion.  The deepest observational survey of the Kuiper Belt to date was reported by \cite{bernstein04} who used the Hubble Space Telescope (HST) to look for faint KBOs, down to $m\simeq29$.  Their results show that the brightness distribution of KBOs flattens 
significantly at magnitudes fainter than $m\simeq24$.  The implication is that the population of small KBOs, down in the size regime of JFCs, is far smaller than that estimated from extrapolation of the size distribution of larger KBOs.  Secondly, there is now a larger observational sample of scattered disk objects (SDOs) with improved orbital parameters;  this can be used to improve the fidelity of the dynamical calculation of the scattered disk's contribution to the JFC population, because the numerical simulations are sensitive to the initial conditions of SDOs.  These new data motivate the present work to re-assess the scattered disk--JFC link.

Shortly before we submitted this paper, \cite{fraser08} reported a new ground-based Kuiper belt survey, with limiting magnitude $\sim26.4$; their results are not inconsistent with those of prior ground-based surveys and also of the survey conducted with the HST \citep{bernstein04}, but they do not find evidence for a break in the size distribution.  However, another recent survey by \citet{fuentes08}, with limiting magnitude $\sim25.7$, does find strong evidence for a break in the KBO size distribution, with a power law index at the faint end of the differential size distribution of $-2.5 \pm 1$, which is slightly steeper than, but still consistent with, the \citet{bernstein04} results.  In this paper, we use the results from \citet{bernstein04} as they still represent the deepest probe into the faint-end size distribution.

We describe briefly in section 2 the previous theoretical models and
observational constraints; these indicate that the scattered disk 
is a viable source reservoir for the JFCs but only with extreme choices 
of parameters permitted by the observational constraints; otherwise, 
there is a large discrepancy between observations and theory. 
We then describe in section 3 our improved model for the 
intrinsic orbital distribution of the scattered disk;
our main improvement is in obtaining a better representation of 
the intrinsic orbital element distribution of the scattered disk 
from the current database of known SDOs (which is much larger than 
when previous modelling studies were done).  
The improved model of the scattered disk's orbital distribution
is used to generate initial conditions for a 4--gigayear numerical 
simulation; in section 4 we present results of this simulation 
in which we calculate anew the flux of JFCs that we expect from 
this source.  In section 5, we compare the simulation results with 
observations of the scattered disk and carefully quantify the 
discrepancy between theory and observation of the origin of JFCs. 
In section 6, we discuss possible resolutions of the discrepancy.
We summarize in section 7.

\section{Previous theoretical and observational studies}\label{s:previousmodels}

The Jupiter family comets are defined as comets with Tisserand parameter in the range 2--3~\citep{levison96}.  The low inclination distribution of the JFCs is much better accounted for by a disk-like source than an isotropic source, leading to the conclusion that the disk-like Kuiper belt is the most likely candidate rather than the nearly isotropic Oort Cloud which is thought to be the source of most other comets~\citep{duncan88, quinn90}.  The connection between the JFCs and the KBOs has been explored in several numerical studies~\citep{holman93, duncan95, levison97, duncan97, morby97, emel04, fernandez04}.

The early studies of \citet{holman93} and \citet{duncan95} examined the role of a population of hypothetical low-inclination and relatively low-eccentricity objects with semi-major axes between about 30 and 50 AU in resupplying the JFCs.  These pioneering studies discovered that weak orbital chaos generated by the long term gravitational perturbations of the giant planets provided dynamical pathways to make such a scenario for the origin of the JFCs quantitatively viable. \citet{holman93} estimated that a population of $5\times10^9$ comet-size objects in the Kuiper belt in the 30--50 AU heliocentric distance range is required to account for the observed population of JFCs.  A more detailed calculation, based on the same physical model, by \citet{levison97} revised the Kuiper belt population estimate to $7\times10^9$.  These models were based on what is now described as the ``classical Kuiper belt'', which is dynamically cold (low-eccentricity, low-inclination orbits).  In a continuation of the \citet{levison97} studies, \citet{duncan97} also found that a small fraction of the classical Kuiper belt objects (CKBOs) could be excited to higher inclinations and eccentricities by close encounters with Neptune to form a persistent population of objects dubbed the `scattered disk'; they reported that this population was a more efficient supplier of JFCs, requiring only $6\times10^8$ objects to be the sole source.  In another model, \citet{morby97} suggested that Plutinos (Kuiper belt objects that are in resonant libration in the 3:2 mean motion resonance with Neptune) are also subject to weak orbital chaos and instabilities on gigayear timescales, and thus could also be a viable source of the JFCs; he estimated a comet-sized Plutino population of $4.5\times10^8$.   

In comparisons of the theoretical population estimates to direct observations of scattered disk objects, it is necessary to define the relevant size range of the objects.  We make the reasonable assumption that the source population for the JFCs must consist of objects at least as large as the JFCs themselves.  \citet{tancredi06} report on the size distribution of JFC nuclei, finding that (i) a power law function with index $-2.7$ best describes the differential size distribution of the population, and (ii) most known JFCs have diameters in the range 1---10 km.  When comparing the population estimates, we are therefore considering objects of this size range and larger.  

Early measurements of the number density and size distributions of the classical belt and the scattered disk found that the two populations contain comparable numbers of objects larger than $\sim100$~km in diameter, and that both have differential size distributions described well by a power law function with index $\sim-4$~\citep{jewitt98,trujillo00}.  Extrapolating these distributions to smaller sizes implies that the CKBOs and the SDOs larger than 1 km in diameter would each number $\sim 4\times10^9$. Because the numerical simulations find the SDOs to be a more efficient source than the CKBOs (i.e., the influx of JFCs per unit source population is higher from the SDOs than from the CKBOs), the extrapolations of the ground-based observations suggest that the SDOs should be the overwhelmingly dominant contributor to the JFC population, as argued in \citet{duncan04}. 

A major question that prevents this from being the final word on the origins of the JFCs is whether the extrapolation of the size distribution of the larger KBOS to smaller sizes is valid.  No surveys to date have been able to actually detect kilometer-sized objects in the Kuiper belt region because they are so faint.  The deepest successful survey of the Kuiper belt thus far was conducted by \citet{bernstein04}, who used the HST to probe $\sim0.02$ deg$^2$ of the sky for KBOs down to a limiting magnitude $m\simeq29$.  (For reference, a brightness magnitude $m\lesssim29$ corresponds to diameter $D\gtrsim 15$~km if we assume a 4\% cometary albedo and a nominal 40 AU heliocentric distance.)  This survey detected only three new faint objects with $m\approx28$ (equivalent to $D\approx25$~km), none of which fall into the category of SDOs; in contrast, the extrapolation from ground-based surveys predicted the detection of $\sim90$ objects given the estimate of the survey's limiting magnitude and detection efficiency.  This paucity of faint objects indicated that the brightness distribution of KBOs flattens greatly at faint magnitudes. 

Using the HST survey data and including data from previous ground-based surveys, \citet{bernstein04} constructed double power law fits to the luminosity functions of a ``classical'' KBO population and an ``excited'' KBO population; their definitions of these two classes were based on only two orbital parameters, the ecliptic inclination, $i$, and the heliocentric distance, $d$, at discovery: the ``classical'' KBOs were those objects with $38 \,\AU < d < 55 \,\AU$ and $i\leq5^\circ$; this definition was adopted to exclude most resonant and scattered objects.  The ``excited'' KBOs were defined as all other objects with $d>25\,\AU$.  These definitions of the ``classical" and ``excited" dynamical classes are similar to but not identical with the CKBOs and SDOs adopted in most of the theoretical literature. We will assume that the excited population of \citet{bernstein04} is the same as our SDO population.  The excited population includes some resonant KBOs, but because there are many more SDOs than resonant KBOs, the error introduced is small.  For all the observed KBOs in the 30\,\AU~to 50\,\AU~heliocentric distance range, Bernstein et al.~found different power law indices at the faint-end and the bright-end of the luminosity functions, which they described with the following function: 
\begin{equation} 
	\Sigma(R) = \Sigma_{23} c [10^{-\alpha_1(R-R_{eq})} + c 10^{-\alpha_2(R-R_{eq})}]^{-1}
\end{equation}
where $\Sigma_{23}$ is the sky density at magnitude $R=23$, $R_{eq}$ is the magnitude at which both terms contribute equally to the sky density, $c$ is a constant equal to $10^{(\alpha_2 - \alpha_1)(R_{eq} - 23)}$, and $\alpha_1$ and $\alpha_2$ are fit parameters.  For the excited class \citet{bernstein04} find that at the faint end, $\alpha_2$ must be less than 0.36 and at the bright end, $\alpha_1$ is between 0.58 and 0.80.  The parameters $\Sigma_{23}$ and $R_{eq}$ are found to be 0.52 and 26.0 respectively.  For the size distribution function, the power law index (given by $5\alpha + 1$) for the ``excited'' class at smaller sizes has a best--fit value of $-1.5$ and a 95\% confidence lower limit of $-2.8$; for $D\gtrsim1$~km, these imply a population of $\sim3\times10^5$ (best--fit estimate) to $\sim2.5\times10^8$ (95\% confidence limit).  For comparison with theoretical models, we note that \citet{duncan97} estimate that $\sim1.4\times10^8$ of the SDOs required to supply the JFCs would be within 30 to 50 AU heliocentric distance at any given time. The \citet{bernstein04} best--fit observational estimate falls short of this theoretical estimate by a factor of $\sim500$; the 95\% confidence estimate is marginally consistent.  Thus, the scattered disk can be considered a viable source of the JFCs only if we choose parameters at the extreme end of the observationally permitted values: i.e., the JFCs are of small size, near $D\approx1$~km, {\it and} the faint-end size distribution is near the steepest allowed by the observations.  Small shifts in any of the parameters could change this agreement.

In theoretical estimates, the key parameter is the rate at which objects escape from the scattered disk.  An examination of the dynamical studies shows that this rate is sensitive to the assumed initial conditions in the numerical simulations. \citet{duncan97} found a fractional escape rate of SDOs, defined as the total fraction of objects leaving the scattered disk divided by the length of the numerical integration, of $2.7\times10^{-10}$yr$^{-1}$ (quoted in \citet{levison06}) based on simulations of hypothetical SDOs generated from low-inclination and low-eccentricity orbits in the 30--50 AU zone; it is unclear how well these represent the real scattered disk.  More recently, two other studies have modelled the orbital evolution of SDOs using initial conditions based on real observed SDO orbits; each of these studies differs slightly in their definition of `SDO'. \citet{emel04} performed numerical simulations of the scattered disk based on 7 observed SDOs; they defined an SDO as one having perihelion distance $q<37$~AU and semi-major axis $a>60$~AU.  They found that the fractional escape rate from the scattered disk is $\sim\!10^{-9}$yr$^{-1}$.  \citet{fernandez04} used the definition $q>30$~AU and $a>50$~AU to select SDOs from the list of known trans-Neptune objects while also including a few with $q<30$~AU; from their numerical simulation, the fractional escape rate of SDOs is found to be $\sim\!10^{-10}$yr$^{-1}$.  Clearly the stability of the modeled scattered disk population is sensitive to the assumed initial conditions. (It is also interesting to note that the upper range of the SDO escape rates found in these studies imply that the present scattered disk is a very small remnant, $<2\%$, of the population $\sim4$~gigayears ago.)  Changes in the initial conditions in a numerical model of the real scattered disk's evolution could change the estimated influx rate of JFCs and push the agreement with the observations more firmly one way or the other.

\section{Modeling the scattered disk}\label{s:modeling}

Here we describe our improved model of the scattered disk.  The improvement is achieved partly by including the newest observations of the scattered disk and partly by accounting for observational biases.  Many new objects have been discovered since the previous dynamical models were published, which provide improved statistics of the SDO orbital distribution.  We also use slightly modified criteria for defining which of the observed objects are SDOs; we will define SDOs to be objects with $q>33$~AU and $a>50$~AU.  The use of $q>33$~AU rather than $q>30$~AU to define SDOs is as suggested by the results of \citet{tiscareno03}, whose simulations of the Centaur population (objects that are transitioning between the Kuiper belt region and the inner solar system) indicate that while there is some overlap between the phase spaces of the Centaurs and the SDOs, the relatively short lived Centaurs rarely cross the $q\approx33$~AU boundary.  This classification possibly excludes a few SDOs with $30<q<33$~AU, but ensures that the dynamically distinct Centaurs are not included in the initial conditions for SDOs.  Our condition on the semi-major axis excludes most classical and resonant KBOs.  

As of February 2007, the Minor Planet Center\footnote{http://www.cfa.harvard.edu/iau/lists/Centaurs.html} listed 80 objects that we classify as SDOs ($q>33$~AU and $a>50$~AU).  These objects are subject to the usual observational biases:  objects that spend most of their time at large heliocentric distances are fainter and therefore less likely to be detected, and high inclination objects are less likely than low inclination objects to be found by ecliptic surveys.  These biases must be accounted for when using observational data to create a model that is representative of the scattered disk.

\subsection{Debiasing $a$ and $e$}\label{ss:ae}

The observational bias in semi-major axis, $a$, and eccentricity, $e$, can be quantified by calculating the fraction of an object's orbit that is spent within the limiting magnitude range of the observational campaign that discovered it.  For objects found in large surveys, the limiting magnitudes are generally well determined.  Of the 80 SDOs used here, 31 were discovered by the Deep Ecliptic Survey and 11 were discovered by the Canada-France-Hawaii Telescope Survey.  The statistics and limiting magnitudes for the Deep Ecliptic Survey and the Canada-France-Hawaii Telescope Survey are described by \citet{elliot05} and \citet{trujillo01cfht} respectively.  For the remaining 38 objects not found by dedicated, well characterized Kuiper belt surveys, an effective limiting magnitude was inferred from the distribution of the objects' visual magnitudes upon discovery.  This limiting magnitude may not be accurate for any individual object, but because the goal of the debiasing procedure is to gain insight to the bulk orbital characteristics of the objects, it is a reasonable approach for this group of objects.  Of these objects, 6 were analyzed separately using the same method, because they all used data from the Sloan Digital Sky Survey.  Because detecting moving objects was not the primary objective of this survey, characteristics for this use of the survey have not been published, but analyzing the objects separately yields a better estimate of the limiting magnitude because they all come from data taken with the same instrument under similar conditions.

To estimate the probability of discovery, each object's orbit was apportioned into fractions spent within certain ranges of heliocentric distances.  The first range of distances corresponds to the range in which the object had effectively a 100\% chance of discovery because it would be brighter than the 100\% detection efficiency of its discovery group.  The second range is the range of distances for which the object had less than a 100\% chance of discovery, but more than a 50\% chance, corresponding to the discovery group's 50\% limiting magnitude.  The fraction of the object's orbit spent outside both these ranges was considered to have a 0\% chance of discovery; we use this simplified scheme because (a) the detection efficiency for any given survey tends to drop off very rapidly with magnitude past the 50\% detection limit (see for example \citet{bernstein04}'s Fig.~2 or  \citet{fuentes08}'s Fig.~1), and (b) the uncertainties in the photometry used to determine the visual magnitudes of the discovered objects are high,  and thus do not warrant a more elaborate scheme.  The probability  of discovery for any object is then the fraction of an orbital period the object spends in the first range plus half the fractional time spent in the second range.  

This probability can be used to appropriately weight each of the known objects to produce the model population.  This weighting is included in the number of times each of the known SDOs is `cloned' to create our model population.  The number of clones is inversely proportional to the detection probability because the detection of objects that spend only a small fraction of their time in an observable range implies that the intrinsic SDO population contains many such objects.  The drawback of this cloning procedure is that it leaves some of the observed objects with only one or a few clones in a $\sim10^3$ test particle simulation.  Because the evolution of any single test particle is not particularly meaningful in a simulation that aims to discover the statistical behavior of the entire population, we included additional clones in those instances where only one or two clones were represented.  At the end of the 4 Gyr integration, the weighting of these additional clones was reduced to reflect the original weighting in the debiasing procedure.  For example, if an observed SDO should only have one clone in the simulation according to the debiasing procedure, but instead ten clones are included, each of the ten clones is counted as one--tenth of one test particle in the analysis of the total simulated SDO population.  This procedure allows us to determine the probable evolution of orbits similar to each of the observed objects, most importantly the probability that they will escape the scattered disk, without needing to create a population of test particles so large as to become computationally prohibitive.

\subsection{Debiasing $i$}\label{ss:i}

For surveys performed near the ecliptic, observational biases against large inclinations can be quantified by the probability that an object with inclination $i$ will be found near zero ecliptic latitude.   This probability is proportional to $1/\sin(i)$.  Debiasing the inclinations of our sample of 80 observed SDOs is not so simple, however, because not all of the surveys were performed near the ecliptic, and many of the objects included were not found by well characterized surveys.  Because of this complication, it was decided that instead of using the individual inclinations to calculate a detection probability for individual objects (as done for $a$ and $e$ in Section~\ref{ss:ae}), the inclinations for the model population would be selected from a distribution based on the observed inclination distribution following the procedure of \citet{brown01}.  We note that this choice is justified only if there is no correlation between inclination and combinations of $a$ and $e$.  For the known SDOs this appears to be a good assumption, so we proceeded with separate debiasing procedures.  First, the inclinations of only the objects found near the ecliptic (in this case within $1^{\circ}$ of the ecliptic) are fit to a half-Gaussian:
\begin{equation}\label{eq:ecliptic}
f_e(i) = C \exp{\frac{-i^2}{2\sigma_1^2}}, \qquad i > 0
\end{equation}
where C and $\sigma_1$ are fit parameters.  The total inclination distribution is then given by the ecliptic distribution (eq.~\ref{eq:ecliptic}) multiplied by $\sin(i)$.  The results of this fitting procedure are shown in Figure~\ref{f:dist}.  This procedure is not ideal because it only uses a subset of the observed population (28 out of 80 objects, in this case) to determine the distribution, and has unquantified biases for objects with large eccentricities as noted by \citet{brown01}.   Despite the shortcomings of this method, the inclination distribution it yields is a physically reasonable one.  We find later, in our simulation results, that the inclination appears to be the least important parameter for predicting which objects will leave the scattered disk on gigayear timescales, suggesting that higher or lower inclinations do not much affect an SDO's orbital stability.

\section{Numerical Simulation}\label{s:results}

\subsection{The model}\label{ss:tp}
The debiasing procedures described in Sections~\ref{ss:ae} and~\ref{ss:i} were used to generate initial conditions for a simulated population of 1849 test particles.  The clones of individual observed objects were given the same eccentricity, longitude of ascending node, and argument of perihelion as the original object.  The semi-major axes of all the clones for a given observed object were evenly spread in a 5\% interval around the observed value; this spread is similar to the observational uncertainties in this orbital parameter\footnote{The 5\% uncertainty is adopted because the Minor Planet Center does not report uncertainties in their fitted orbital elements.   67 out of the 80 objects used here have been observed at multiple oppositions, so their orbit fits are well constrained.  The orbits of the remaining objects are less well determined, but their inclusion or exclusion did not statistically alter the distributions of the test population, so they were included.}.  The distributions of $a$, $e$, and $q$ for the observed population and the simulated population are shown in Figure~\ref{f:dist}.  The inclinations were randomly selected from the best fit inclination distribution, and the mean anomaly for each test particle was randomly selected from 0 to 2$\pi$.  

Our debiased population can be compared to the debiased population of \citet{disisto07} who modeled the scattered disk and Centaur populations.  The authors employ a similar debiasing procedure to our own for the semi-major axis distribution, with the addition that they superimpose a power law distribution of semi-major axes after applying the correction factor for heliocentric distance.  Their result is a semimajor axis distribution that is peaked in the $40-70$ AU region, whereas ours starts at $50$ AU and is relatively flat in the $50-90$ AU range.  While the classical disk does seem to follow a power law in heliocentric distance, there are indications that the scattered population's heliocentric distance distribution is much more flat in the range of $40-60$ AU \citep{kavelaars08}.  \citet{disisto07} also calculate a debiased inclination distribution; their distribution has a peak that is sharper than our distribution and located at a lower inclination, but it is roughly consistent within our $1-\sigma$ error bars.

The orbits of the simulated SDOs (treated as massless test particles), were integrated under the gravitational influence of the Sun and the four outer planets using a mixed variable symplectic integrator based on the algorithm of \cite{wh91}; we carried out the orbital integrations for 4 Gyr, using a step size of 1 year.  The initial conditions for the test particles were as specified above, and the initial conditions for the planets were taken from the JPL Horizons service\footnote{Found at http://ssd.jpl.nasa.gov/?horizons}~\citep{jplhorizons} for 9 February 2007.  We stopped integration of an individual particle if it reached a distance of 10,000 AU or if it approached within a Hill radius of one of the planets; we consider such particles to have `left' the SDO population.

\subsection{Results}\label{ss:results}
Our main result from this simulation is the escape rate of SDOs.  We find that 42\% of the objects left the scattered disk in the 4 Gyr length of the integration.  The average rate at which SDOs leave the scattered disk is then estimated as the fraction leaving divided by the length of the integration; this is found to be $1.2\times10^{-10}$yr$^{-1}$.  To check the consistency of this rate over time, the fraction of objects leaving the scattered disk relative to the number of objects remaining was calculated at several times in the integration.  When we do this, we find that the rate leaving during the first Gyr of the simulation is $2\times10^{-10}$yr$^{-1}$, twice as large as the rate of $1\times10^{-10}$yr$^{-1}$ during the 1--4 Gyr period.  The reason for this can be seen in Figure~\ref{f:removaltimes} which plots a histogram of the times at which particles `left' the scattered disk.   The distribution peaks during the first Gyr then levels off for the remainder of the integration.  We interpret this as an imperfection of the debiasing procedure, specifically the tendency for a few low-detection-probability, unstable SDOs in the observed sample to dominate the outflux in the first Gyr.  This is discussed further in section~\ref{ss:analysis}.

As mentioned in Section~\ref{ss:i}, there does not appear to be a strong correlation between the initial inclination of a particle and the probability that it will leave the scattered disk over gigayear timescales.  The same holds true for $a$ and $e$.  There is, however, a tendency for objects with smaller initial perihelia to escape preferentially compared to objects with large initial perihelia.  This is shown in Figure~\ref{f:q_escape} which plots the initial perihelion distance distribution of the SDOs that escaped. We find that few objects with initial perihelion distances larger than 40 AU escape, although several of the objects that did escape evolved to perihelia as large as 50--to--60~AU before escaping.

\subsection{Analysis}\label{ss:analysis}
A simple property we demand of our debiased model is that it be nearly stationary in time, i.e., that the model distribution not favor a special time; this provides basic confidence in our debiasing procedure.  To determine if our model distribution satisfies this, we perform a Kolmogorov-Smirnov test (adapted from \citet{press92}) to compare the distributions of the orbital elements (semi-major axis, eccentricity, perihelion distance and inclination), $a$, $e$, $q$, and $i$, at various times in the simulation.  The test measures the absolute difference in the cumulative distributions of two samples of any given orbital element and translates this into a probability, which we will call the `KS probability', that the two samples are drawn from the same parent distribution.  This is accomplished by comparing the absolute difference between the samples' cumulative distributions to the distribution of differences expected for two samples that are drawn from one distribution.  If the debiasing procedure does create a model distribution that is representative of the scattered disk, we expect the K-S test to yield a high KS probability for the distributions of each orbital element compared at different times in the simulation.

We find that the KS probability is nearly 100\% for the inclination distributions compared at the beginning and the end of the simulation, indicating that the inclination distribution obtained in the debiasing procedure is stationary on 4 Gyr timescales.  The eccentricity distributions at 0 Gyr and at 4 Gyr give a KS probability of 81\%, whereas the distributions at 1 Gyr and at 4 Gyr yield nearly 100\%.  The KS probabilities for the semi-major axis and perihelion distance distributions at 0 Gyr and at 4 Gyr are only 43\% and 5\%, respectively.  The probability for the semimajor axis reaches nearly 100\% when we consider the distributions at 1 Gyr and 4 Gyr.  For the perihelion distance, the KS probability increases to 56\% when we consider the distributions at 1 Gyr and 4 Gyr, and it is nearly 100\% for the comparison between 2 Gyr and 4 Gyr.  The improvement in the KS probability when the first gigayear of the simulation is excluded indicates that the initial population has a transient subset, but the distribution stabilizes after 1 Gyr and remains stationary for the remainder of the simulation.  This can be seen most clearly in the evolution of the perihelion distribution, which is shown in Figure~\ref{f:qdist_time}.  The objects with a very low probability of discovery that initially dominate parts of the distributions of $q$ and $a$ due to the debiasing procedure are either lost from the scattered disk or evolve to more stable regions of phase space during the first gigayear.  This supports our hypothesis that the imperfections of the debiasing procedure, specifically the effects of a few low-probability detections of SDOs that are unstable, can explain the difference in the rate of objects leaving the scattered disk during the first gigayear and the remainder of the simulation.

With the above considerations, we conservatively report the escape rate of SDOs as $(1-2)\times10^{-10}$yr$^{-1}$ where the former is the average over the last 1 Gyr and the latter is the average over the first 1 Gyr of the simulation.  A small source of error in this estimate is introduced by our assumption that all the particles that had a close encounter with a planet will leave the scattered disk after their first such encounter. To estimate the magnitude of this error, we note that \citet{duncan97} found that only $\sim5\%$ of trans-Neptunian objects will persist on stable trans-Neptunian orbits after their first Neptune encounter; therefore, the error in our estimate due to this assumption is $\sim5\%$, well within the range reported here.

For comparison, our estimate of the SDO outflux rate is nearly an order of magnitude larger than the $4\times10^{-11}$yr$^{-1}$ found by \citet{duncan95} for the Classical Kuiper belt, confirming that the scattered disk is a less stable population than the Classical belt.  The rate for the scattered disk that we find here is an order of magnitude smaller than the rate found by \citet{emel04}; a likely explanation for this difference is that those authors considered only initial orbits with perihelia interior to 37 AU, a sample that is overall less stable than one that includes the full range of perihelia representative of the scattered disk.  When only objects with initial $q<37$ AU in our simulation are considered, the fractional escape rate is $(1.6-3.5)\times10^{-10}$yr$^{-1}$, which is a factor of $\sim\!3$ smaller than the \citet{emel04} estimate.  The rest of the discrepancy might be explained by the difference in the number of observed SDOs contributing to each sample;  \citet{emel04} based their population on 7 real SDOs, while our sample of real SDOs with $q<37$ AU is 43.  Our estimate is also smaller than that of \citet{disisto07}, who report that SDOs enter the Centaur population at a fractional rate of $5\times10^{-10}$ yr$^{-1}$; the discrepancy here is likely due to the difference between their debiased semi-major axis distribution, which is sharply peaked at smaller semi-major axes, and ours, which is more flat (see Section~\ref{ss:tp}).  Our estimate is closer in agreement to the rate $2.7\times10^{-10}$yr$^{-1}$ obtained from the simulations of \citet{duncan97} (as reported by \citet{levison06}); the latter was based on only 33 scattered disk--type particles, compared with our sample of 80.  Our estimate also agrees with the estimate of \citet{fernandez04}; however their work was based on 76 observed objects, only 49 of which meet the criteria for SDOs as given in this paper.

\subsection{Population of the scattered disk: theoretical estimate}\label{ss:SDOpop} 
The rate at which objects leave the scattered disk can be used to estimate the total number of SDOs by making the assumption that the scattered disk is in steady state with the JFC population.  Previous studies~\citep{levison97,fernandez04} have consistently shown that 30\% of objects will evolve to JFC orbits after their first encounter with Neptune.  This means that the fractional rate at which SDOs enter the JFC population is $(3-7)\times10^{-11}$yr$^{-1}$.  In steady state, this influx of SDOs into the JFCs must balance the outflux of JFCs. In order to estimate the latter, we need to know about the current visible JFC population, the visible lifetime of the comets, and the dynamical lifetimes of JFC orbits.  We will take the definition of visibility as a JFC with $q<2.5$~AU (following \citet{levison97}), because such comets are usually bright and come close enough to the Earth that they should represent an observationally complete population.  There are currently about 250 JFCs that satisfy this criterion.  Next, the population of non-active cometary nuclei (so-called `dormant' comets) must be accounted for.  Based on simple physical models of volatile depletion in comets, \citet{jewitt04} estimates the ratio of dormant to active comets to be $\sim\!2$ for JFC--sized objects.  This brings the total number of JFCs with $q<2.5$~AU up to 750.  To then estimate the intrinsic total JFC population, we need to know how much time a typical JFC spends with $q<2.5$~AU.  In a dynamical study of known comets, \citet{levison94} found that JFCs spend about 7\% of their time in this visible range.  The final estimate of the JFC population is then 10,000 objects.  \citet{levison94} also found that JFCs have a median dynamical lifetime of $3\times10^{5}$ years. Therefore, the loss rate of the JFC population is $\sim\!3.3\times10^{-2}$ comets per year.  For the scattered disk to balance this loss, there must be a total of (0.6--1.1)$\times10^{9}$ SDOs.  

In the following, we refer to this estimate of the scattered disk population as the ``theoretical estimate", because it is based on a theoretical calculation of the outflux rate of SDOs; of course, we also use observations of the JFCs in obtaining the actual numbers.  The largest source of uncertainty in this theoretical estimate of the SDO population is the adopted value of the total JFC population.  If we assume that the observed population of active JFCs larger than 1 km diameter is complete for objects with $q<2.5$ AU, the uncertainty in the total population is primarily in the estimate of the dormant--JFC population and in the estimate of the JFC dynamical lifetime.  We adopted a value of 2 for the ratio of dormant to active comets, based on simple physical models by \citet{jewitt04}.  However, dynamical considerations led \citet{levison97} to conclude that the ratio could be as high as 7.  Resolving these conflicting estimates will require a detailed understanding of the physical processes that cause comets to become dormant, and/or better observational constraints on the population of dormant comets.  Preliminary studies of what are thought to be dormant JFCs in the near Earth object population suggest that the ratio is between 1 and 3 \citep{fernandez06}, supporting our adopted value.  The dynamical lifetime of JFCs is the other major source of uncertainty because of the limitations of the \citet{levison94} models; their numerical integrations were too short to sample the full range of JFC lifetimes, and the integrated population was based on the known, and therefore, observationally biased sample of JFCs.  The error introduced by this estimate is unknown; the dynamical lifetime of the JFCs should be reassessed in future work. 

\section{Comparison to observations of the scattered disk}\label{s:sd_obs}

How does our theoretical estimate for the population of the scattered disk compare with estimates from direct observations of the SDOs?  To make this comparison we need to consider the population of JFC-size SDOs.  The JFC population has been determined to be dominated by objects of 1--10 km in diameter~\citep{tancredi06,lowry08}; such objects in the heliocentric distance range 30--50 AU would have magnitude $m\sim30$ and fainter (assuming the average albedo of $0.04$ for cometary nuclei).  Surveys for faint objects in the outer solar system have yet to achieve limiting magnitudes in this range, but the size distributions determined from detections of brighter objects can be extrapolated to give an estimate of how many comet--sized objects exist in the scattered disk.  At present, the size distribution determined by \citet{bernstein04} probes deepest into the small end of the trans-Neptunian population in the 30 to 50 AU distance range.  The limiting magnitude of this survey is $m=29$, which just approaches the range of interest for JFC-size objects.  The authors find a turnover in the scattered disk size distribution near $m\simeq25$, which implies far fewer small objects than estimated from an extrapolation of the single power law that describes the larger objects.

To compare the results of our simulations to the \citet{bernstein04} size distribution, we must account for the heliocentric distance and ecliptic latitude range of the observations.  The reported size distribution is for objects in the 30--50 AU zone observed close to the ecliptic, so it is necessary to know what percentage of our simulated SDO population is in this range at any given time.  To do this, we calculated the test particles' heliocentric distance distribution averaged over the last 100 Myr of the integration;  the result is shown in Figure~\ref{f:helio}: on average, 15\% of our particles can be found in the 30--50 AU heliocentric distance range. This corresponds to a population of $(0.8-1.7)\times10^{8}$.  The ecliptic latitudes of the objects were also calculated over the last 100 Myr, and on average 15\% of the simulated SDOs are within $\pm3^\circ$ of the ecliptic, which corresponds to the latitude range of most KBO surveys.  Figure~\ref{f:size_dist} shows the resulting ecliptic sky density for the simulated SDOs compared to the \citet{bernstein04} cumulative size distribution.  The objects in the source population for the JFCs must be at least as large as the JFCs themselves, so the observed distribution needs to overlap the theoretical population estimate in the 1--10 km diameter range, i.e.~in the magnitude range 30--35.  From Figure~\ref{f:size_dist}, we see that even for the most favorable extrapolation of the observations, the theoretical estimate from our simulation just barely agrees in this range.  For the best--fit size distribution the discrepancy with theory is more than two orders of magnitude.  

\section{Possible reasons for the discrepancy}\label{s:discrepancy}

\subsection{Incompleteness of observed JFC population}

In obtaining our theoretical estimate of the scattered disk population, we use the current observational estimate of the JFC population, we assume a value of the ratio of dormant to active comets, and we use the current estimate of JFC dynamical lifetimes. We have used conservative assumptions for the observational estimate and the ratio of dormant to active comets; any revisions to either of these factors would cause the theoretical estimate to increase, not decrease, thus worsening the discrepancy.  The dynamical lifetime estimate is based on numerical modeling and its error is undetermined, so it is unclear if this would increase or decrease the discrepancy, but the estimate would need to be too short by an order of magnitude or more to erase the gap between our theoretical prediction for the scattered disk and the current observations.  Therefore, other explanations for the discrepancy must be sought. 

\subsection{Cometary albedos}\label{ss:albedos}
A source of uncertainty that might contribute to the discrepancy is the conversion of magnitude to size when comparing population estimates .  Doing this requires an assumption about the albedos of SDOs.  The standard practice has been to assume an average albedo of 0.04.  This value is based on the measured albedos of many short and long period comets~\citep{lamy04}, and it was adopted here wherever a conversion between magnitude and size was required.  However, this albedo might not apply uniformly to CKBOs and SDOs; \citet{grundy05} report on the wide range of albedos measured for KBOs and suggest that 0.1 would be a more reasonable assumption than the canonical 0.04.  \citet{stansberry08} similarly report a wide spread of albedos for KBOs and Centaurs.  They also report that larger objects have a tendency toward higher albedos, and that there are hints of a trend amongst the Centaurs for objects with smaller perihelia to have lower albedos (although these are still mostly higher than 0.04).  

If these higher albedos were used to convert between magnitude and sizes, our definition of JFC-sized SDOs would be shifted toward brighter magnitudes.  In Fig.~\ref{f:size_dist}, the ``theoretical estimate'' would likewise shift to brighter magnitudes.  This increases the range of magnitudes over which the theoretical estimate and the observations could overlap. It also makes the discrepancy larger. So we must conclude that the discrepancy here is in fact a lower limit, and that there are no reasonable assumptions about albedos that will reconcile the best--fit observations with the theoretical estimate.

\subsection{Change in SDO size distribution beyond the observational limit}\label{ss:sfd} 
The current limiting magnitude of the deepest observational survey is $m\simeq29$ \citep{bernstein04}, but JFC-sized objects at heliocentric distances of $\sim\!40$ AU have magnitudes in the range 30--35.  A steepening of the power law slope past the observational limit might offer a possible resolution to the discrepancy between the observations and our theoretical estimate.  If we take the best-fit sky density at $m=29$ from \citet{bernstein04}, and assume that the sky density required by the simulations corresponds to $m=35$, the differential size distribution defined by these two end points has a power law index $-3.3$.  Pegging the theoretically required sky density to smaller magnitude values leads to a steeper power law index; for example, using the value $m=32$, which is the middle of the magnitude range for JFC-sized SDOs, the required power law index is $\sim\!-6$.  The former value is significantly beyond the range of the observational uncertainty for the faint-end SDO size distribution, and the latter falls far outside it.  Values of the power law index smaller than $-4$ are unphysical if such a size distribution extends to very small sizes because it would imply an infinite total mass.

While it is just barely possible to resolve the discrepancy by appealing to an upturn of the size distribution of the SDOs at the very faint-end, it should be noted that this would indicate that the slope for the putative JFC source population is steeper than that measured for the JFCs themselves in the 1---10 km range.  \citet{tancredi06} report the steepest power law size distribution for the JFCs, with an index of $-2.7$.  Many other groups report slopes much shallower than this, with power law indices ranging from $-1.4$ to $-1.9$ (see \citet{lowry08} for a recent review).  It is possible that the size distribution of the JFCs does not match that of the source population, but any physical process that results in a shallower size distribution for the JFCs would have to preferentially remove small SDOs as they become JFCs; this means the required total source population for the JFCs would need to be even larger than the estimates presented here.  Clearly, an upturn of the SDO size distribution beyond current observational limits is a possible, but still problematic way to resolve the discrepancy. 

\subsection{Non-steady state}\label{ss:steady-state}
We might also wonder if the steady state assumption is correct, that is, that the JFC population is in steady state with its source, or if perhaps the presently observed JFC population is a large fluctuation above its long term average. We note that the assumption of steady state is reasonable for the present solar system, because any very unstable populations in the outer solar system have likely long been depleted, making any unusual increase in the flux of objects into the JFC population unlikely near the present epoch.  
Still, we can test this unlikely possibility by considering a closely related group of objects, the Centaurs.  

Centaurs are a dynamically distinct transitional population which represent an intermediate dynamical stage between the JFCs and their trans-Neptune precursors. If the JFCs are in steady state with the SDOs, then the Centaurs should be as well, and we can predict the Centaur sky density in this case.  Adopting the fractional escape rate of SDOs found in our simulations, a median dynamical lifetime of 9 Myr for Centaurs \citep{tiscareno03}, and the observed SDO size distribution from \citet{bernstein04}, we calculate that the total population of Centaurs larger than $D \sim 50$ km should be $\sim200$--300. Assuming a nominal heliocentric distance 20 AU (based on the known Centaurs) and 10\% albedo (based on the measured albedos of $D\sim100$~km KBOs), a $D=50$~km Centaur has brightness $m\simeq 22.5$. Therefore, this Centaur population estimate is equivalent to an ecliptic sky density of (2--3)$\times 10^{-2}$deg$^{-2}$ at magnitude $m\simeq 22.5$, as indicated in Fig.~\ref{f:size_dist}.  A much larger Centaur population would be indicative of a large fluctuation above the steady state production rate from the SDOs.

We can also calculate the Centaur population in the JFC size-range ($D\gtrsim1$~km) that would be in steady state with the observed JFCs.  As discussed in Section~\ref{ss:SDOpop}, the JFCs need to be replaced at a rate of $3.3 \times 10^{-2}$yr$^{-1}$ to maintain the current population.  \citet{tiscareno03} find that approximately one third of the Centaurs enter the JFC population. Taking these together, we obtain the fractional loss rate of Centaurs: 0.1 per year.  Then using the Centaur median dynamical lifetime of 9 Myr, we obtain that $\sim 10^6$ Centaurs larger than $D \sim 1$ km must exist in order to balance the loss of JFCs.  This estimate is two orders of magnitude smaller than that of \citet{disisto07}, but their estimate of the Centaur population relies on a steady state assumption with SDOs based on a single power law size distribution for the SDOs down to 1 km sizes.   There are now several lines of evidence for a break in the KBO size distribution at sizes much larger than this \citep{bernstein04,fuentes08,lowry08}, so the \citet{disisto07} Centaur estimate is likely too large.  \citet{horner04} also make an estimate of the Centaur population based on dynamical considerations, finding a total number $4.4\times10^4$ larger than $D\sim1$ km.  However, the authors use outdated estimates of the required influx of new JFCs to calculate the number of Centaurs required to maintain steady state; this likely accounts for the discrepancy with our estimate.  \citet{sheppard00} estimate the total Centaur population larger than $D\gtrsim1$ km to be $\sim10^7$ based on the then current population of observed Centaurs.  This is closer in agreement with our estimate, but the small number of observed objects and the large uncertainties in the assumed size distribution make the \citet{sheppard00} estimate quite uncertain.

These predictions for the Centaur population based on a Centaur-SDO steady-state or a JFC-Centaur steady state cannot be tested at present, as there are no observational estimates of the Centaur population from a well characterized survey, but are testable with future observations.  

\subsection{Other sources}\label{ss:other-sources}
A possible resolution of the discrepancy would be a different or additional source for the JFCs.  \citet{duncan95} and \citet{levison97} show that weak instabilities allow classical belt objects to enter the JFC population, and \citet{morby97} showed that Plutinos (resonant KBOs) can also be a source of the JFCs.  However, both the CKBOs and the resonant KBOs as major sources of JFCs have fallen out of favor in recent years \citep{levison06,duncan04}.   
Current estimates of the required populations of these two dynamical classes are a factor of 10 or more larger than the population of this class inferred from observations~\citep{bernstein04}, and thus do not help resolve the problem readily.  We note, however, that the faint-end slope of the CKBOs is steeper than that of the SDOs and is similar to that of the observed JFCs. 
We also note that the fractional escape rate of the CKBOs is $\sim$~(0.3--0.5)$\times10^{-10}$yr$^{-1}$~\citep{duncan95} and that for the Plutinos is $0.5\times10^{-10}$yr$^{-1}$~\citep{morby97}, values that are only a factor 2--5 less than that for the SDOs found here.  Thus, we think that there is a need to reassess the CKBOs and the resonant KBOs as sources of the JFCs.  This is beyond the scope of the present paper, but we hope to do this in future work.  

Another possible source region might be the Jupiter Trojan asteroids.  \citet{jewitt00} estimate that there are $1.6\times10^5$ Trojans larger than 1 km.  Dynamical studies of known Trojans by \citet{tsiganis05} suggest that 14\% of these have orbits that are not stable over the age of the solar system.  Even if every object that left the Trojan population became a JFC, this would only contribute $\sim6\times10^{-6}$ comets per year to the JFC population, far short of the $3.3\times10^{-2}$ comets per year needed to account for the observed JFCs in steady state.  Therefore, the Trojans are not likely to be a dominant contributor to the JFC population.

\subsection{Breakup of SDOs?}\label{ss:tidal}

A plausible physical mechanism that might explain the apparent deficit of SDOs compared to JFCs is breakup of the SDOs into multiple fragments at some point during their transport from the trans-Neptune source region to the inner solar system.  
The splitting of comets appears to be not uncommon, as there is observational evidence for at least 40 such events in very recent history (reviewed recently by \citet{boehnhardt04}).  Some of the observed splitting events occurred near perihelion for sungrazing comets, indicating tidal breakup into many large fragments, and other splitting events have occurred at larger heliocentric distances. In the latter cases, the splitting is more akin to peeling off outer layers that became unstable due to thermal or rotational stresses~\citep{sekanina97}.  The breakup of comet Shoemaker-Levy 9 (SL9) after a close encounter with Jupiter showed that comets can also be tidally split as a result of planetary encounters.  These tidal splitting events in which the parent comet splits into several large fragments definitely happen, but their frequency will determine whether they can resolve the discrepancy between the scattered disk source and the observed JFC population. Is such a hypothesis quantitatively viable?

Models of SL9's encounter with Jupiter reveal that the original comet is best described as a nearly strengthless rubble pile prior to the breakup event~\citep{asphaug96,walsh03}.  Investigations of the tidal disruption of near-Earth asteroids using a rubble pile model~\citep{richardson98,walsh06} show that there are three basic types of breakups:  SL9--type {\it catastrophic} breakups where the largest surviving fragment contains less than 50\% of the original mass, {\it moderate} breakups where the largest fragment has 50-90\% of the mass, and {\it mild} breakups where less than 10\% of the original mass is lost.  These results are consistent with the observed breakup of SL9 into $\sim20$ fragments, and the breakup of comet 16P/Brooks2 into three major components after an encounter with Jupiter in 1886~\citep{sekanina97}.  The frequency of these different types of events is highly uncertain because the type of breakup is very sensitive to the comet's closest approach distance to the larger body~\citep{asphaug96}. 

Crater chains on the Galilean satellites may offer some clues to the historical frequency of SL9 type events.  \citet{schenk96} analyze 11 crater chains on Ganymede and Callisto, each consisting of 6 to 25 craters, and conclude that they are most likely the result of impacts immediately following a SL9-type tidal breakup.  Using this cratering record, the authors estimate that catastrophic breakups occur roughly once every 300 years in the Jupiter system.  Because objects on their way in from the transneptunian region suffer many close encounters with all the outer planets \citep{tiscareno03}, some may very well experience a break-up event.

Is there evidence in the short-period comet population for the prevalence of splitting events creating multiple comets?  As of 2004, ten of the 160 known short-period comets were known to have split, although some of these events were more akin to mass shedding than splitting events that produce persistent secondary nuclei~\citep{boehnhardt04}.  \citet{tancredi00a} report the results of using Lyapunov indicators to look for break-up families in the JFC population; they conclude that groups of 10 or more comets sharing a common break-up event are not prevalent in the 123 JFCs used in the investigation.  So while there are observations of splitting events and even pairs of comets that share a common origin \citep{boehnhardt04}, there is no direct evidence of larger sets of JFCs sharing a single progenitor.  This could be a selection effect if the secondary nuclei produced by break-ups have shorter fade times than the primary nucleus, thereby becoming very difficult to observe.  It could also be due to the small numbers of comets that have been extensively studied.  It also may be more difficult to trace back to break-ups that occurred earlier in a comet's history during encounters with planets other than Jupiter.   

A quantitative estimate of the effect of tidal disruption on the supply rate of JFCs requires a careful calculation of the distribution of close encounter distances and knowledge of the physical strength properties of the objects; we do not attempt such a calculation here.  For the purpose of illustration, we can estimate the fraction of comets that are tidally disrupted as at least as large as the fraction of comets that will impact a giant planet during their lifetimes; in order to break up, a comet must come within a few planetary radii of the giant planet \citep{asphaug96}, so this is a reasonable approximation.  \citet{levison00} estimate that 2\% of ecliptic comets will impact a giant planet.  Taking the \citet{bernstein04} population estimate for the scattered disk, the rate at which SDOs become JFCs, and the rate at which the JFCs need to be resupplied, the 2\% of SDOs that experience tidal disruption would need to break into roughly $100-1000$ fragments to account for the discrepancy.  A population that could offer clues about the efficiency of break-up events, and thus if the rate and extent of tidal disruption comes close to the required limit estimated above, is the Centaur population.  During their dynamical lifetime, Centaurs suffer many close encounters with the four outer planets.  \citet{tiscareno03} analyzed the orbital evolution of 53 observed Centaurs and reported a total of $\sim 8000$ close planetary encounters over the dynamical lifetimes of the 53 objects.  They also report that $(4\pm2)\%$ of their objects impacted a planet, which is in agreement with the \citet{levison00} estimate.  If tidal break-up occurs for a significant fraction of the Centaurs, then there are two effects on the Centaur population: their numbers would be greater than expected from the steady state dynamical models, and the size distribution of Centaurs would be steeper than the SDO source (having relatively larger number of small objects). 

At present, the data on the sizes and total population of Centaurs is very limited due to a lack of well-characterized surveys sensitive to Centaur detection.  The power law index of the size distribution is estimated to be about $-4$ for objects larger than $D\sim100$~km (similar to KBOs), but the small-end size distribution has not been constrained~\citep{sheppard00}.  Deeper observations of the Centaurs can test the SDO break-up hypothesis by examining population counts and the size distribution.

\section{Summary}\label{s:summary}

\begin{figure}[htbp]
\begin{center}
\includegraphics[width=5.5in]{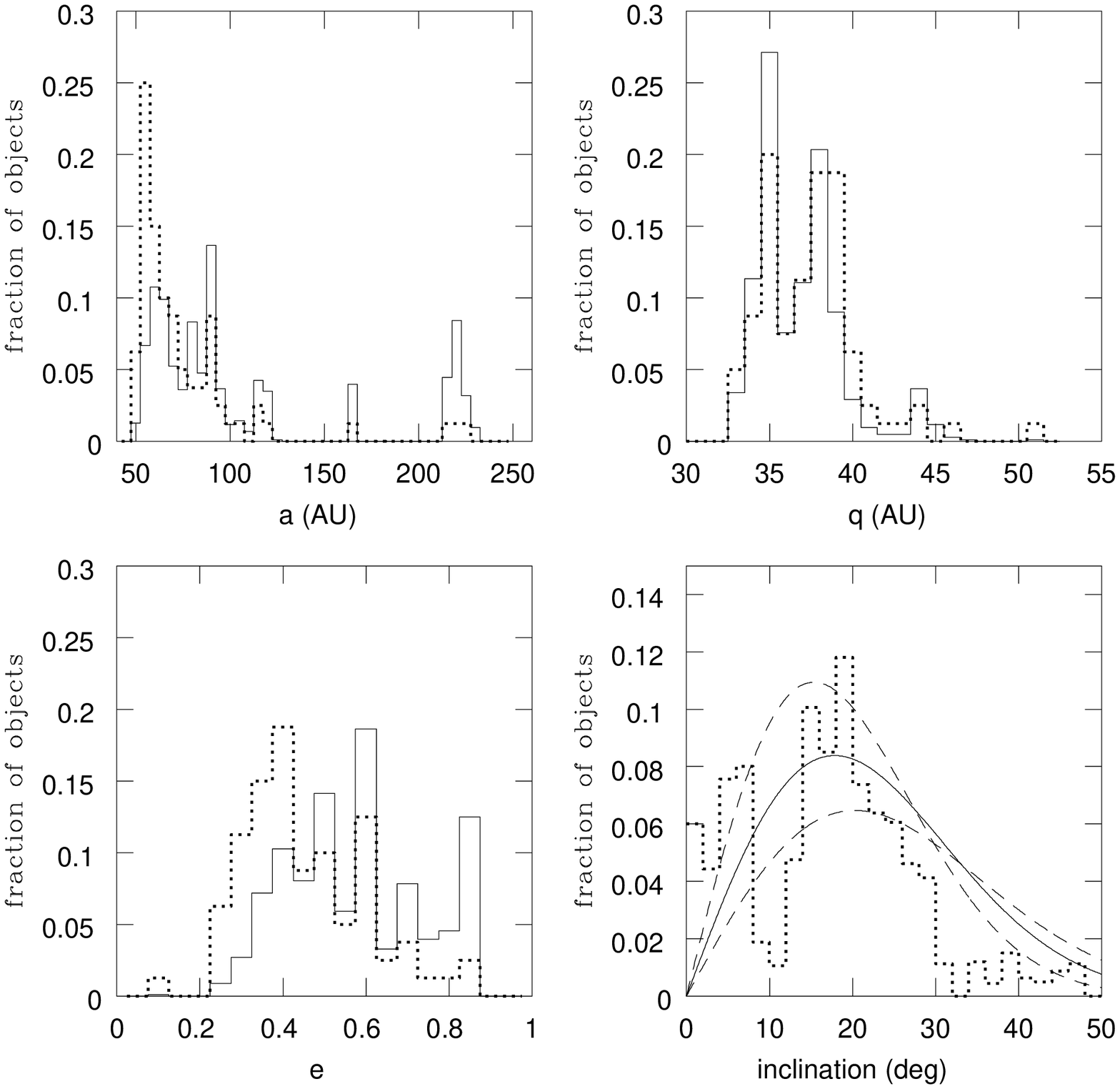}
\caption{{ The observed distributions of a, e, and q (dotted lines) compared to the distributions for the debiased model population (solid lines). The plot for the inclination shows the inclinations of all observed SDOs (histogram) compared to the modeled distribution (solid line); the 1$\sigma$ errors (dashed lines) were derived from the fit to the ecliptic inclination distribution.  (See Section \ref{ss:i}). }}
\label{f:dist}
\end{center}
\end{figure}

\begin{figure}[htbp]
\begin{center}
\includegraphics[width=5.5in]{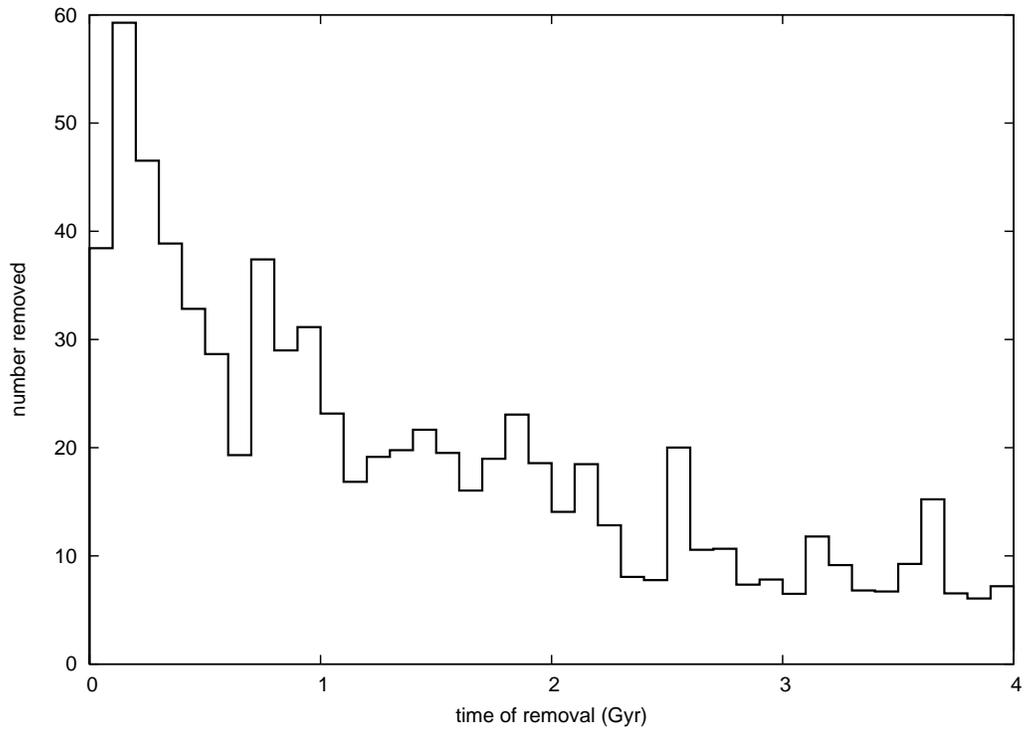}
\caption{{ Histogram of the times at which particles in the simulation were removed, either due to a planetary encounter or by evolving to heliocentric distance $\gtrsim10,000\AU$.}}
\label{f:removaltimes}
\end{center}
\end{figure}

\begin{figure}[htbp]
\begin{center}
\includegraphics[width=5.5in]{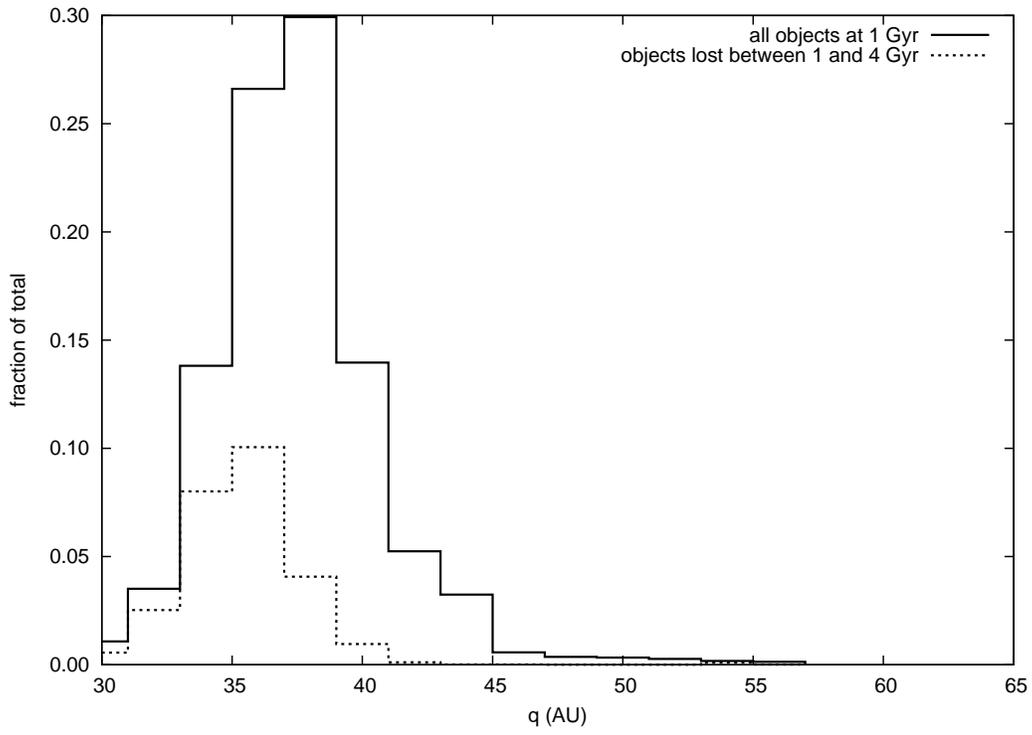}
\caption{{ The distribution of perihelia at 1 Gyr for the model particle population compared to the perihelia of particles that escaped from the scattered disk between 1 and 4 Gyr.}}
\label{f:q_escape}
\end{center}
\end{figure}

\begin{figure}[htbp]
\begin{center}
\includegraphics[width=5.5in]{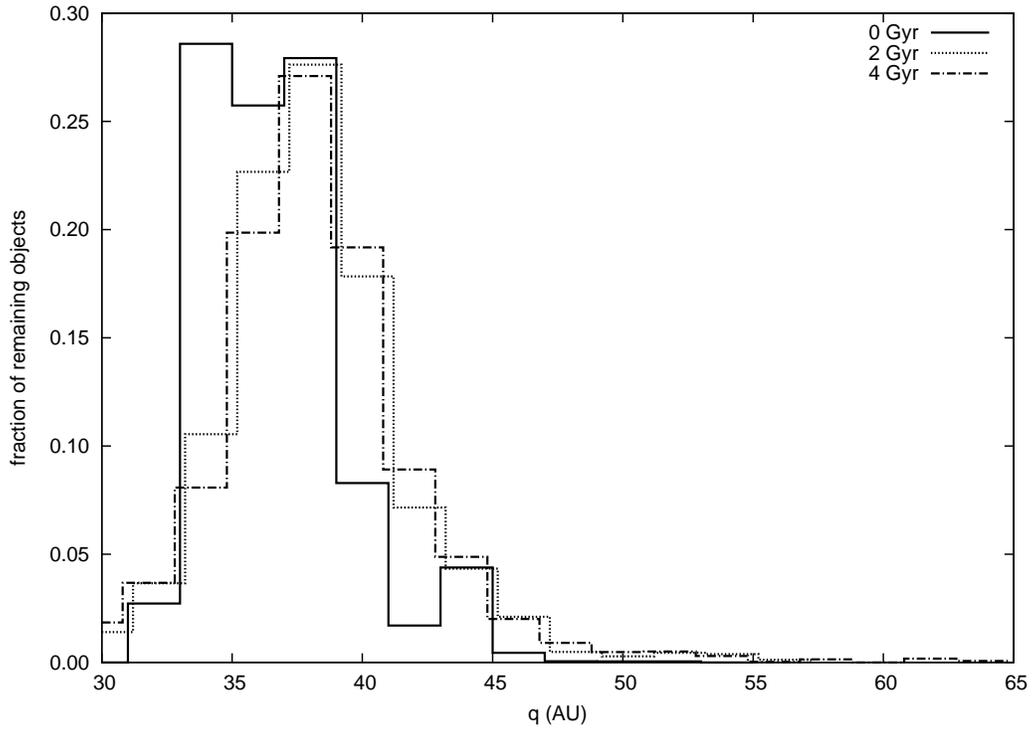}
\caption{{ Snapshots of the distribution of perihelion distance at several epochs during the simulation. The distribution at 0 Gyr differs significantly from the distribution at 4 Gyr, but the distributions at 2 and 4 Gyr are statistically similar.}}
\label{f:qdist_time}
\end{center}
\end{figure}

\begin{figure}[htbp]
\begin{center}
\includegraphics[width=5.5in]{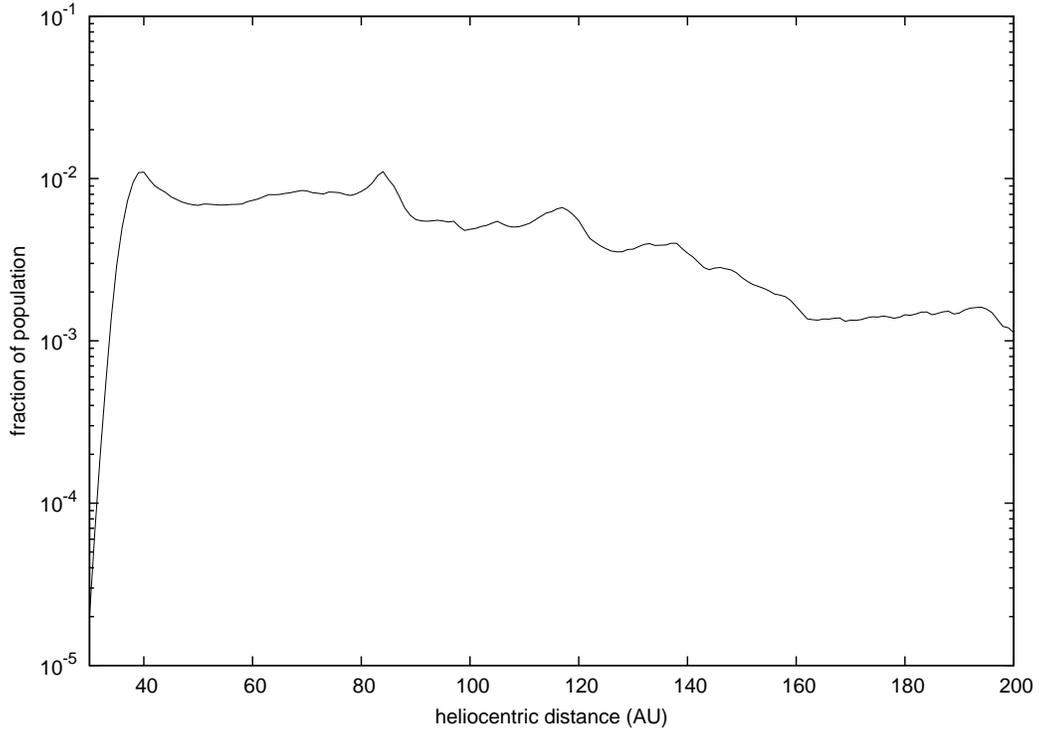}
\caption{{ The heliocentric distance distribution of all the remaining test particles averaged over the last 100 Myr of the integration.  Note the relatively flat distribution in the 40--90 AU range in our model. This is in agreement with the observationally-derived distribution in~\citet{kavelaars08}. }}
\label{f:helio}
\end{center}
\end{figure}

We have quantitatively assessed the viability of the scattered disk as a source reservoir of the short period Jupiter family comets.  To do this, we constructed a ``debiased" model of the orbital distribution of the scattered disk (see Figure \ref{f:dist}), and numerically simulated the orbital evolution of $\sim1800$ test particles representative of SDOs for 4 billion years to obtain the rate at which they encounter Neptune.  Our calculation of this rate is in fairly good agreement with the dynamical models of \citet{duncan97}, \citet{fernandez04}, and \citet{levison06}, but is at variance with that of \citet{emel04}.  Using this rate together with the results of previous studies on (i) the dynamics of ecliptic comets \citep{levison94}, (ii) the physical evolution of short period comets \citep{jewitt04}, and (iii) data on the observed JFCs \citep{tancredi06}, we estimate the SDO population required to supply the JFCs in steady state.  We compare this with the results of the deepest observational surveys of the trans-Neptunian population thus far.  Our results and conclusions are as follows.

\begin{figure}[htbp]
\begin{center}
\includegraphics[width=5.5in]{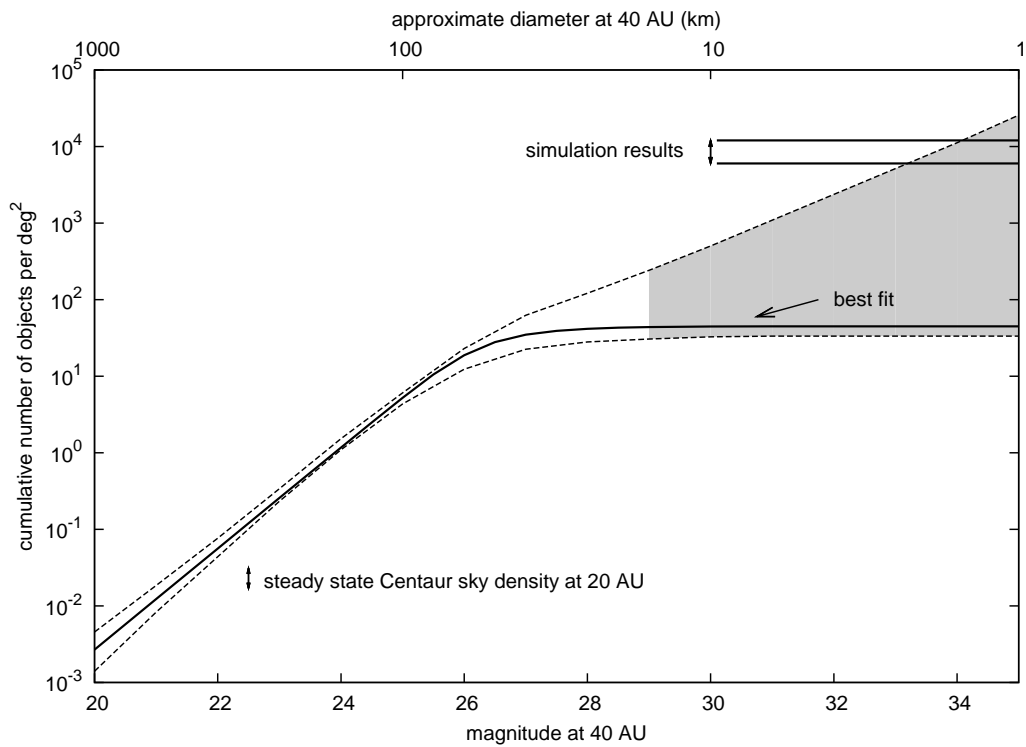}
\caption{{ The cumulative size distribution of scattered disk objects from \citet{bernstein04} compared to the population estimate from this work (horizontal line at top right).  The solid line is the best fit distribution, with the dashed lines showing the possible ranges. The gray shaded region indicates the extrapolation past the observational limit.  The vertical arrows in the lower portion of the graph indicate the predicted sky density for the Centaurs if they are in steady state with the observed SDO population of sizes larger than $D\sim 50$ km ($m=26$ at 40 AU and $m=22.5$ at 20 AU).}}
\label{f:size_dist}
\end{center}
\end{figure}

\begin{enumerate}
\item  Our theoretical calculations find that a population of at least (0.8--1.7)$\times10^{8}$ comet-sized (diameter $D> 1$ km) scattered disk objects in the 30--50 AU heliocentric distance range is necessary in order to supply the observed population of Jupiter family comets in steady state on gigayear timescales. (The ecliptic sky density of such a population would be (0.6--1.2)$\times10^{4}$~deg$^{-2}$, Fig.~\ref{f:size_dist}.).  In contrast, the present observational estimate of this population is $3\times10^{5}$ to $2\times10^{8}$; here the lower and upper numbers are based on the best-fit and on the 95\% confidence limit, respectively, of the observed faint-end size distribution \citep{bernstein04}.  The best-fit observational number falls short of the theoretical requirement by more than two orders of magnitude; the 95\% confidence limit is marginally consistent with theory. 

\item  Consideration of uncertainties in the dynamical estimates makes the discrepancy worse.  We find the fractional escape rate of SDOs to be (1--2)$\times10^{-10}$yr$^{-1}$; we adopted a conservative estimate of the uncertainty in our calculation of this rate: the lower end of the reported range is likely more representative (Section~\ref{s:results}).  Our theoretical estimate of the total SDO population depends on observational estimates of the current total JFC population; uncertainties in the latter imply that our SDO population estimate is a lower limit (Section~\ref{s:sd_obs}).  Errors associated with the uncertain values of cometary albedos (used to convert size to magnitude) only increase the discrepancy.  These considerations suggest that our theoretical estimate of the required scattered disk population is a lower limit, and therefore the uncertainties in the dynamical estimates cannot resolve the discrepancy. 

\item A change in the size distribution function of the SDOs just beyond the current observational limit is a possible but problematic way to resolve the discrepancy. Such a change in the SDO size distribution would need to be present implausibly close to, but just beyond the present observational limit of $m\simeq29$; the power law index of the SDO size distribution at the small-size end would need to be $\leq-3.3$, which is steeper than the observational uncertainty.  It is also also much steeper than that for the JFCs themselves.

\item  Alternative sources --- such as the classical Kuiper belt, the resonant KBOs, or the Jupiter Trojans --- may contribute to the JFC population, as there are dynamical pathways available.  Given current understanding of their populations and their long term dynamics, these do not appear to be viable sole sources of the JFCs: the observed populations fall short of the required contribution by one--to--three orders of magnitude.  However, we note two points about the CKBOs: this population has a faint-end size distribution which may be significantly steeper than that of the SDOs \citep{bernstein04}, and its fractional escape rate (reported in \cite{duncan95}) is only a factor $\sim\!3$ less than what we found for the SDOs in our simulations.  Therefore, in future work, it is worth re-assessing the CKBOs as a possible source of JFCs.  A similar case may exist for reassessment of the resonant KBO source.

\item  An improbable solution for the discrepancy is that the observed JFCs represent a large fluctuation above their long term average population.  Such an explanation is difficult to understand physically, as it would require a recent heavy influx of JFCs, presumably from an unstable pocket within the trans-Neptunian population.  Such a sub-population would need to survive for several gigayears, then become destabilized only within the last $\sim10^7$ years to provide the observed JFC population.  At present, we know of no dynamical mechanism that can accomplish this.  However, we offer predictions for the population of Centaurs that could test this unlikely explanation (Section~\ref{ss:steady-state}).

\item  A possible solution for the discrepancy is that fragmentation of larger objects during the orbital evolution from the trans-Neptune region to the inner solar system is important in increasing the population of objects that eventually become JFCs.  The Centaurs, the transitional population between trans-Neptunian objects and the JFCs, undergo frequent close encounters with the giant planets  \citep{tiscareno03}.  It is estimated that $\sim2\%$ of Centaurs will encounter a planet at a small enough distance to induce tidal disruption;  these 2\% of Centaurs would need to break into 100--1000 fragments to account for the discrepancy (Section~\ref{ss:tidal}).  There is some evidence for breakup families within the JFC population \citep{boehnhardt04}, so this explanation is not implausible, but it does require more fragmentation than has been seen in current models of tidal disruption \citep{asphaug96}.  Probing the size distribution of the Centaur population could test this hypothesis and indicate how important tidal disruption is for supplying the JFCs. 

\item  As a final point, we note that the escape rate of SDOs that we found implies that the current scattered disk population represents $40-66\%$ of the population from $\sim\!4$ Gyr ago.  This means that, over the history of the solar system, the depletion of this putative source of the JFCs has been relatively modest, and not as high as some previous studies had implied.  This estimate can help constrain theoretical models of the origin of the scattered disk.

\end{enumerate}

\acknowledgements
We thank Luke Dones for a careful and helpful review.  This research was supported by a grant from NASA's Outer Planets Research program.

\bibliographystyle{apj}
\bibliography{ms}

\clearpage

\end{document}